\documentclass{ws-procs975x65}
\usepackage{graphicx}

\def\beq{\begin{equation}}
\def\eeq{\end{equation}}

\begin{document}

\title{Early-time cosmological solutions in scalar-Gauss-Bonnet theory}
\author{Panagiota Kanti}

\address{Department of Physics,University of Ioannina, Ioannina GR-45110, Greece\\
Department of Physics, National and Kapodistrian University of Athens, Athens, Greece\\
E-mail: pkanti@cc.uoi.gr}

\begin{abstract}
We consider a gravitational theory that contains the Einstein term, a scalar field and the
quadratic Gauss-Bonnet term. We focus on the early-universe dynamics, and demonstrate
that the Ricci scalar does not affect the cosmological solutions at early times, when the
curvature is strong. We then consider a pure scalar-GB theory with a quadratic coupling
function: for a negative coupling parameter, we obtain solutions that contain always an
inflationary, de Sitter phase, while for a positive coupling function, we find instead expanding
singularity-free solutions. 
\end{abstract}


\bodymatter


\section{Introduction}

The generalised gravitational theories that contain, apart from the Einstein-Hilbert term,
additional gravitational terms or extra fields, have been intensively studied over the last
decades during the quest for the ultimate gravitational theory that would be valid both
at large and small energy scales. These theories contain, among others, the heterotic 
superstring effective theory \cite{Zwiebach, Gross, Tseytlin} or the Lovelock theory
\cite{Lovelock}. The quadratic Gauss-Bonnet term is present in both of the aforementioned
theories and its implications for gravity and cosmology have been extensively studied
\cite{Antoniadis, KRT, KMRTW, Torii, KKK, Nojiri, Amendola:2007ni, Koivisto:2006xf,
Carter:2005fu,Leith:2007bu}.

Being a topological invariant in four dimensions, the Gauss-Bonnet (GB) term must be
coupled to a field, usually a scalar field, to remain in the theory. Therefore, in this work
we consider a 4-dimensional theory that contains the Ricci scalar, the GB term
and a non-minimally coupled scalar field. We will focus on the dynamics of the early
universe, and argue that during that era the presence of the Ricci scalar adds nothing
to the dynamics of the cosmological solution. For the case of a quadratic coupling
function between the scalar field and the GB term, we demonstrate that this coupled
system leads to cosmological solutions with interesting characteristics that are classified
by the sign of the coupling function constant.


\section{The Theoretical Framework}

We will focus on a generalised gravitational theory that contains the usual
Einstein term -- i.e. the Ricci scalar $R$, a scalar field $\phi$ and the higher-derivative
Gauss-Bonnet term $R^2_{\rm GB}$. The action functional of this string-inspired theory reads
\begin{eqnarray}  
{\cal S}=\int d^4x \,\sqrt{-g} \left[\frac{R}{2}-\frac{(\nabla \phi)^2}{2} +
\frac{1}{8}\,f(\phi) R^2_{\rm GB}\right].
\label{action}
\end{eqnarray} 
The scalar field is coupled non-minimally to gravity via a general coupling function
$f(\phi)$ to the quadratic Gauss-Bonnet term, that is defined as
\beq
R^2_{\rm GB} = R_{\mu\nu\rho\sigma} R^{\mu\nu\rho\sigma}
- 4 R_{\mu\nu} R^{\mu\nu} + R^2\,.
\eeq
We will also assume that the aforementioned scalar field is the dominant agent 
at the early universe and all other forms of matter/energy can be ignored during
that time.

If we vary the action (\ref{action}) with respect to the scalar field $\phi$ and the
metric tensor $g_{\mu\nu}$, we obtain the scalar and gravitational field equations, respectively.
Assuming that the line-element is described by the Friedmann-Lema\^itre-Robertson-Walker
form
\beq
ds^2 = -dt^2 +a^2(t)\left[\frac{dr^2}{1-kr^2}+r^2\,(d\theta^2 + 
\sin^2\theta\,d\varphi^2)\right],
\label{metric}
\eeq
with a scale factor $a(t)$ and spatial curvature $k=0, \pm 1$, the field equations take
the explicit form
\begin{align}
\ddot \phi+3 H \dot \phi-3f' \bigl(H^2+\frac{k}{a^2}\bigr)
(H^2+\dot H) &=0\,, \label{dilaton-eq-0} \\
3(1+H \dot f)\bigl(H^2 +\frac{k}{a^2}\bigr) &= \frac{\dot \phi^2}{2}\,,
\label{Ein-1}\\
2(1+H\dot f) (H^2+\dot H)+(1+\ddot f) \bigl(H^2+
\frac{k}{a^2}\bigr) &=-\frac{\dot \phi^2}{2}\,.\label{Ein-2}
\end{align}
In the above, we have used the expression of the Gauss-Bonnet term for the
line-element (\ref{metric}), that is
\beq
R^2_{GB}=24\,\bigl(H^2+\frac{k}{a^2}\bigr) (H^2+\dot H)\,,
\label{GB}
\eeq
where $H \equiv \dot a/a$ is the Hubble parameter and the dot denotes
derivative with respect to time. 

One cannot help noticing that, for a constant coupling function $f(\phi)$, the GB term
being a topological invariant completely disappears from the field equations. On the
other hand, when present in the theory, it provides a non-trivial potential for the scalar
field that is otherwise forced to adopt a constant configuration. A question that readily
arises is whether the presence, or absence, of the Ricci scalar really modifies the
dynamics of the ensuing cosmological solutions at very early times. We address this
question in the following section.


\section{The Complete Theory with a Linear Coupling}

In this section, we will assume that the function $f(\phi)$ is a linear function of the 
field $\phi$, i.e. $f(\phi)=\lambda \phi$, where $\lambda$ is a coupling constant. Then,
the scalar equation (\ref{dilaton-eq-0}) can be easily integrated once with respect to
time to yield the relation
\beq
\dot \phi=\frac{C}{a^3} + \frac{\lambda \dot a\,(3k+\dot a^2)}{a^3}\,,
\label{rel-phi-adot}
\eeq
with $C$ an integration constant. Clearly, the effect of the GB term to the scalar-field
dynamics is encoded in the second term of the above expression, that is proportional
to $\lambda$. Henceforth, we will focus on this term and thus we set, for simplicity, $C=0$.

\begin{figure}[t]
\begin{center}
\includegraphics[width = 0.53 \textwidth] {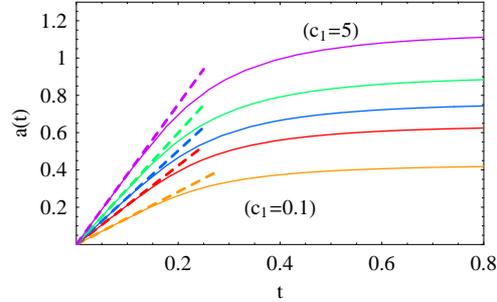}
    \caption{The scale factor $a(t)$ in terms of time, for a linear coupling function and
$\lambda=0.1$, and for the values $c_1=0.1, 0.5,1,2,5$ (from bottom to top).}
   \label{Linear}
\end{center}
\end{figure}

If we rearrange Eqs. (\ref{dilaton-eq-0})-(\ref{Ein-2}), and use Eq. (\ref{rel-phi-adot})
to replace any remaining $\dot \phi$'s (for further details on this, see
 [\refcite{KGD-long}]), we arrive at the constraint equation
\beq
3(k+\dot a^2)^2\left[4 +\lambda^2\,\frac{3 \ddot a}{a^3}\,(k+\dot a^2)\right]
+\lambda^2\,\frac{\ddot a {\dot a}^2}{a^3}\,(3 k +\dot a^2)^2=0.
\eeq
The above equation can be explicitly integrated twice only for $k=0$, in which case
it leads to the solution 
\beq
a(t)\,F\left[\frac{1}{4},\frac{1}{4},\frac{5}{4};\frac{3 a^4(t)}{c_1}\right]=
\left(\frac{2c_1}{5\lambda^2}\right)^{1/4}(t+t_0)\,.
\label{solan1C0}
\eeq
In the above, $F(a,b,c;x)$ stands for the hypergeometric function and $c_1$ is a constant.
The general behaviour of $a(t)$ is depicted in Fig. \ref{Linear}. In the limit $a \rightarrow 0$,
the above reduces to a linear relation between the scale factor and the time coordinate. 

We will now focus on the early-time limit from the beginning and ignore the presence
of the Ricci scalar. Then, a similar rearrangement of the field equations leads directly
to the constraint $\ddot a=0$ and to the linear solution $a(t)=A t+B$ -- these solutions
are also shown in Fig. \ref{Linear}. Therefore, as
expected, in the strong-curvature regime and when the quadratic GB term is present,
the linear term adds nothing to the dynamics of the cosmological solution and thus
it may be altogether ignored. The same conclusion follows by considering the solution
for the scalar field.


\section{The Scalar-GB Theory with a Quadratic Coupling}

Unfortunately, a similar analysis for the case of a quadratic coupling function cannot
be performed due to the complexity of the corresponding field equations. However,
as the coupling function $f(\phi)$ determines the weight of the GB term in the theory,
we expect that its explicit form merely determines the point in time where the GB term
begins to dominate over the linear Ricci term. Thus, choosing appropriately the time
regime, we can always consider a pure scalar-GB theory and its dynamics at the 
early universe.

Therefore, focusing on the case of a quadratic coupling function, i.e. $f(\phi)=\lambda \phi^2$,
and repeating the previous analysis, we arrive now at the constraint  \cite{KGD-short, KGD-long}
\beq
(k+5{\dot a}^2)\,\ddot a \,a^3+
24 \lambda\,{\dot a}^2\,(k+{\dot a}^2)^2=0\,.
\label{Eq1and2-v2}
\eeq
The above equation does not contain $\phi(t)$, and, for the case again of a flat universe
($k=0$), it can be integrated once to give
\beq
\frac{5}{2 {\dot a}^2}=C_1 -\frac{12 \lambda}{a^2}\,.
\label{general}
\eeq
Depending on the values of the integration constant $C_1$ and the Gauss-Bonnet coupling
parameter $\lambda$, we may obtain a variety of cosmological solutions with interesting
characteristics. Here, we present a sample of them classified by the sign of $\lambda$. 


\subsection{The case with $\lambda < 0$}

If the coupling constant $\lambda$ is negative, then the integration constant $C_1$ can
have every possible value. Starting with the case of $C_1=0$, a simple integration of
Eq. (\ref{general}) yields the solution \cite{KGD-short, KGD-long}: 
\beq
a(t)=a_0\,\exp\biggl(\pm\sqrt{\frac{5}{24 |\lambda|}}\,t\biggr).
\label{sol-C10}
\eeq
If we choose the positive sign, the above solution describes an inflationary,
de Sitter-type cosmological solution. The scalar field, whose expression may
easily be found through Eq. (\ref{Ein-1}), is also given by a similar exponential,
but decaying, expression. Both quantities evolve faster with time the smaller
the coupling constant $\lambda$ is. This so-called GB inflation \cite{KGD-short}
is supported only by the GB coupling of the scalar field and needs no additional
potential with particular features as in the more traditional models 
\cite{Guth, Linde, Starobinsky}. The necessary number of e-foldings can also
follow without introducing trans-planckian field values \cite{KGD-short}.

\begin{figure}[t]
\includegraphics[width = 0.51 \textwidth] {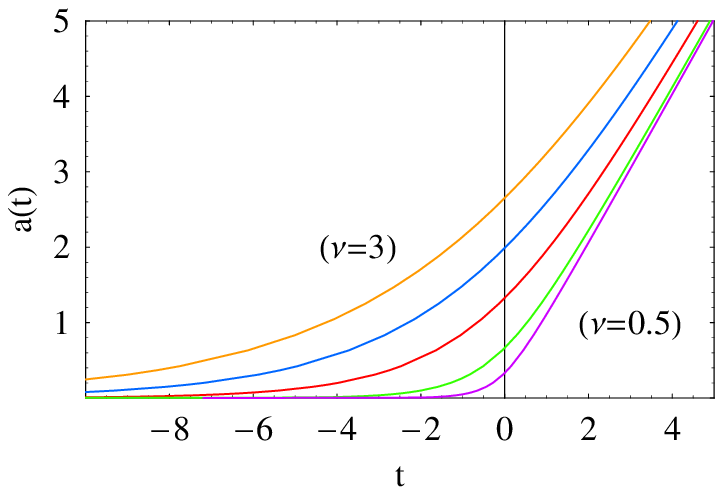} \hspace*{-0.6cm}
\includegraphics[width = 0.51 \textwidth] {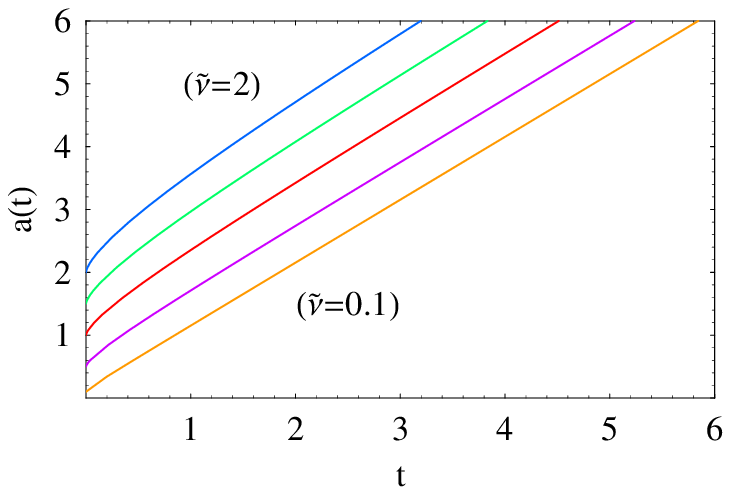}
    \caption{The scale factor $a(t)$ versus time for the cosmological
    solutions with: (a) $\lambda<0$, $C_1>0$ and $\nu=0.5, 1, 1.5, 2, 3$,
    and (b) with $\lambda>0$, $C_1>0$ and $\tilde \nu=0.1,0.5,1,1.5,2$.}
   \label{Fig2}
\end{figure}

A more interesting family of solutions arises in the case where $C_1$ is positive.
In that case, Eq. (\ref{general}) upon integration leads to the relation
\beq
\sqrt{a^2+\nu^2} + \nu \ln\left(\frac{\sqrt{a^2+ \nu^2}-\nu}{a} \right)
= \pm \sqrt{\frac{5}{2 C_1}}\,(t+t_0)\,,
\label{sol-lam-neg}
\eeq
where $\nu^2 \equiv 12 |\lambda|/C_1$. The behaviour of $a(t)$ in terms of time,
for the solution with the (+)-sign, is depicted in Fig. \ref{Fig2}(a): the scale factor
is expanding with time,
with a faster pace at early times and a smaller one later. Indeed, in the limit 
$a \rightarrow 0$, we recover the pure de Sitter solution (\ref{sol-C10}) found
above, while for $a^2 \gg \nu^2$, we obtain a Milne-type linearly expanding solution.
Thus, these solutions accommodate an early inflationary phase with a natural exit
mechanism at later times. The scalar field can be expressed in terms of the
scale factor and is found to follow a decreasing pattern between two constant
values \cite{KGD-short, KGD-long}. 


\subsection{The case with $\lambda>0$}

If the coupling parameter $\lambda$ is positive, then $C_1$ is allowed to take
only positive values, too. This time, Eq. (\ref{general}) leads, upon integration
again, to the relation \cite{KGD-long}
\beq
\sqrt{a^2-\tilde \nu^2} -\tilde \nu \arccos\left(\frac{\tilde \nu}{a}\right)=
\pm \sqrt{\frac{5}{2 C_1}}\,(t+t_0)\,.
\label{sol-lam-pos}
\eeq
where now $\tilde \nu^2 \equiv 12 \lambda/C_1$. The mathematical consistence of
the above relation demands that $a^2 \geq \tilde \nu^2$. As a result, $\tilde \nu$ is
the smallest allowed value of the scale factor in this model, and the corresponding
solutions are thus singularity-free. The profile of the solution, for the (+)-sign, is
depicted in Fig. \ref{Fig2}(b). For large values of the scale factor, the expansion
becomes linear; for small values of $a(t)$ though, we find the asymptotic solution
\cite{KGD-long}
\beq
a(t) \simeq \tilde \nu\,[1+ (A t+B)^{2/3}]\,,
\eeq
which reveals its regular behaviour for any finite values of the time-coordinate
and the role of $\tilde \nu$ as the lower bound of the scale factor of the universe.
The scalar field for this class of solutions starts from a zero value and increases
towards an asymptotic constant value.

A similar analysis for a general coupling function $f(\phi)=\lambda \phi^n$, where
$n$ is an integer number, reveals \cite{KGD-long} that expanding, singularity-free
solutions arise only for the case of $n=2$ studied above. We consider this
not to be a coincidence since the quadratic coupling function falls into a general
class of functions that have similar characteristics with the heterotic string effective
coupling function \cite{KRT}-- singularity-free solutions were numerically shown to
exist in the context of the latter theory \cite{Antoniadis} as well as in the theory with a
quadratic coupling function \cite{KRT}, and here we have demonstrated their
existence analytically even in the absence of the Ricci scalar.

%

\section{Conclusions}

In the context of the Einstein-scalar-Gauss-Bonnet theory we have
demonstrated that the Einstein term of the action, the Ricci scalar, can safely
be ignored at the very-early-time limit when the spacetime curvature is strong.
By considering then a pure scalar-GB theory with a quadratic coupling function,
we have derived a number of cosmological solutions with attractive characteristics.
For negative values of the coupling parameter, 
cosmological solutions that are pure de Sitter or solutions
with a de Sitter phase in their past and a linearly expanding phase later on arise.
For positive values of the coupling constant, a different class of solutions is
found, that contains expanding solutions with no Big-Bang singularity. All the
solutions were derived analytically and, in our opinion, point towards some hidden,
fundamental importance of the quadratic coupling function.

\section*{Acknowledgments}

I would like to thank my collaborators Naresh Dadhich and Radouane Gannouji
for an enjoyable and fruitful collaboration.
This research has been co-financed by the European
Union (European Social Fund - ESF) and Greek national funds through the
Operational Program ``Education and Lifelong Learning'' of the National
Strategic Reference Framework (NSRF) - Research Funding Program: 
``THALIS. Investing in the society of knowledge through the European
Social Fund''. Part of this work was supported by the COST Action MP1210
``The String Theory Universe''.


\end{document}